\documentclass[sigconf]{acmart}

\usepackage{multirow}

\setcopyright{cc}
\setcctype{by}
\copyrightyear{2026}
\acmYear{2026}
\acmDOI{10.1145/3837062.3838901}
\acmConference[MODELS Companion 2026]{ACM/IEEE 29th International Conference on Model Driven Engineering Languages and Systems}{October 04--09, 2026}{Málaga, Spain}
\acmBooktitle{ACM/IEEE 29th International Conference on Model Driven Engineering Languages and Systems (MODELS Companion 2026), October 04--09, 2026, Málaga, Spain}
\acmISBN{979-8-4007-2903-4/2026/10}

\acmSubmissionID{15975.59}

\begin{document}

\title{ModBench: A Pipeline for Building Modelica Benchmark Datasets Mined from Library Repositories}

\author{Masoud Sadrnezhaad}
\orcid{0000-0002-3437-1681}
\email{masoud.sadrnezhaad@liu.se}
\affiliation{%
  \institution{Link\"oping University}
  \city{Link\"oping}
  \country{Sweden}}

\author{Martin Sj\"olund}
\orcid{0000-0001-7638-0108}
\email{martin.sjolund@liu.se}
\affiliation{%
  \institution{Link\"oping University}
  \city{Link\"oping}
  \country{Sweden}}

\author{Adrian Pop}
\orcid{0000-0003-0091-1181}
\email{adrian.pop@liu.se}
\affiliation{%
  \institution{Link\"oping University}
  \city{Link\"oping}
  \country{Sweden}}

\author{Jos\'e Antonio Hern\'andez L\'opez}
\orcid{0000-0003-2439-2136}
\email{joseantonio.hernandez6@um.es}
\affiliation{%
  \institution{University of Murcia}
  \city{Murcia}
  \country{Spain}}

\author{Torvald M\r{a}rtensson}
\orcid{0000-0003-1438-0182}
\email{torvald.martensson@liu.se}
\affiliation{%
  \institution{Link\"oping University}
  \city{Link\"oping}
  \country{Sweden}}

\author{D\'aniel Varr\'o}
\orcid{0000-0002-8790-252X}
\email{daniel.varro@liu.se}
\affiliation{%
  \institution{Link\"oping University}
  \city{Link\"oping}
  \country{Sweden}}

\renewcommand{\shortauthors}{Sadrnezhaad et al.}

\begin{abstract}
Research on equation-based cyber-physical systems modeling languages, such as Modelica, is constrained by the lack of curated benchmark datasets.
This limits empirical insight into the evolution and development of models.
We address this gap with \emph{ModBench}, a pipeline that mines Git repositories of Modelica libraries to produce benchmark datasets of model snapshots.
The pipeline (1)~filters repository commits to retain human-authored, Modelica-relevant revisions; (2)~extracts simulation-eligible classes; and (3)~builds canonical representations of Modelica classes.
For empirical validation, we applied ModBench to the Modelica Standard Library (MSL) and report the resulting dataset, spanning the full commit history (since Modelica language v3), with 85562 distinct class snapshots, and links enabling traceability to original models and Git metadata.
The dataset, its API, and the data generation pipeline are publicly available to support future research on model evolution analysis, compiler testing, and automated model repair or generation.
\end{abstract}




\maketitle


\newcommand{\MbToolOmc}{v1.28.0}

\newcommand{\MbFilterInputCommits}{10{,}139}
\newcommand{\MbFilterExclBranches}{223}
\newcommand{\MbFilterAfterBranches}{9{,}916}
\newcommand{\MbFilterExclPreVThree}{1{,}334}
\newcommand{\MbFilterAfterPreVThree}{8{,}582}
\newcommand{\MbFilterExclBot}{88}
\newcommand{\MbFilterAfterBot}{8{,}494}
\newcommand{\MbFilterExclNoMo}{1{,}134}
\newcommand{\MbFilterRetained}{7{,}360}
\newcommand{\MbFilterRetainedPct}{72.6}             

\newcommand{\MbListDistinctClassNames}{9{,}067}
\newcommand{\MbListAllClassVersions}{38{,}903{,}069}
\newcommand{\MbListNonExperimentVersions}{36{,}320{,}196}
\newcommand{\MbListNonExperimentPct}{93.4}
\newcommand{\MbListExperimentVersions}{2{,}582{,}873}
\newcommand{\MbListExperimentPct}{6.6}
\newcommand{\MbListMaxExperimentPerCommit}{535}

\newcommand{\MbGrowthAllClassMean}{5{,}546}

\newcommand{\MbGrowthAllClassMax}{6{,}373}

\newcommand{\MbGrowthExperimentMean}{368}

\newcommand{\MbGrowthExperimentRatioMean}{6.5}

\newcommand{\MbCanonCommitsWithExperiment}{7{,}015}
\newcommand{\MbCanonCommitsWithExperimentPct}{95.3}
\newcommand{\MbCanonProducedVersions}{2{,}505{,}835}
\newcommand{\MbCanonProducedPct}{97.0}
\newcommand{\MbCanonFailures}{77{,}038}
\newcommand{\MbCanonFailuresPct}{3.0}

\newcommand{\MbStorePipelineDbSize}{1.2\,GB}
\newcommand{\MbStoreClassEnumDbSize}{12\,GB}
\newcommand{\MbStoreCanonicalDirSize}{199\,GB}

\newcommand{\MbFailOperatorRestriction}{61{,}960}
\newcommand{\MbFailOperatorRestrictionPct}{80.4}
\newcommand{\MbFailMissingClass}{7{,}274}
\newcommand{\MbFailMissingClassPct}{9.4}
\newcommand{\MbFailRedeclare}{5{,}825}
\newcommand{\MbFailRedeclarePct}{7.6}
\newcommand{\MbFailMissingBase}{1{,}023}
\newcommand{\MbFailMissingBasePct}{1.3}
\newcommand{\MbFailTimeout}{675}
\newcommand{\MbFailTimeoutPct}{0.9}
\newcommand{\MbFailPackageLoad}{281}
\newcommand{\MbFailPackageLoadPct}{0.4}

\section{Introduction}

Empirical research on cyber-physical systems (CPS) modeling languages is hindered by the lack of curated benchmark datasets~\cite{boll2026data_desert,lopez2022modelset,DBLP:journals/sosym/BollBAKV21}. Modelica — an acausal, equation-based language widely used for multi-domain physical system simulation — is no exception. The challenge is not a shortage of raw data: public Modelica libraries are plentiful~\cite{tiller2014impact}. Rather, it is the absence of curated, normalized artifacts that capture the evolution of real-world models in a form suitable for empirical analysis.
Without such benchmarks, researchers gain limited insight into how models evolve, how development processes unfold, how regressions are introduced and resolved, and how modeling practices develop~\cite{DBLP:journals/sosym/BollBAKV21}.

Mainstream software engineering research has long benefited from established benchmark corpora for languages such as Java (e.g., Defects4J~\cite{DBLP:conf/issta/JustJE14}).
Modelica, however, is an object-oriented, acausal, declarative language expressing system dynamics via differential-algebraic equations, properties which call for dedicated datasets.

Moreover, producing a runnable simulation from a Modelica class is a multi-stage process carried out by a Modelica compiler such as OpenModelica (OM), and model semantics emerge only after flattening equations.
Consequently, file-level mining (of \texttt{.mo} sources), which is the strategy underlying mainstream software-repository mining pipelines~\cite{DBLP:journals/tmlr/KocetkovLALMJMF23}, is insufficient to identify simulation-eligible units or to compare them across commits in the Modelica context.
Therefore, existing software engineering pipelines cannot be directly reused for Modelica libraries to construct meaningful benchmarks. These pipelines require dedicated, compiler-assisted extraction that operates on elaborated classes rather than raw text.

We present \emph{ModBench}, which provides three research artifacts: (1)~a reusable \emph{generation pipeline} that mines Modelica library histories by filtering repository commits, listing Modelica classes, and constructing canonical Modelica source files for simulation-eligible classes; (2)~the \emph{ModBench-MSL corpus}, a dataset produced by running the pipeline on the MSL; and (3)~a \emph{query API} layer that provides uniform read access over the generated databases and canonical files.
The pipeline is the primary methodological contribution, designed for reuse across other Modelica libraries, while the MSL corpus validates it on a real-world Modelica project at scale.
To assess transferability beyond the MSL, we also tested the pipeline on the DLR thermofluid stream library.

Altogether, ModBench offers four benefits.
\emph{Executability:} canonical files are compiler-validated and self-contained, so each retained snapshot can be loaded and simulated without recovering its original repository context.
\emph{Realism:} the corpus is grounded in high-quality, human-authored models that reflect real engineering decisions rather than synthetic examples.
It also retains non-compilable variants as failure records that are valuable for error analysis and compiler diagnostics.
\emph{Traceability:} each canonical snapshot is linked to its source repository, commit, and Modelica class name, supporting reproducibility and cross-referencing with issues and pull requests.
\emph{Evolutionary history:} the corpus spans the MSL development history from Modelica language version~3 onward, with millions of class--commit pairs, enabling studies of regression patterns, refactoring trends, and model maturity over time.

\section{Modelica Background}
\label{sec:background-modelica}

\noindent\emph{Modelica classes.}
Modelica classes represent components, connectors, records, functions, or complete models, and describe systems primarily using equations.
Classes can inherit from and extend other classes. They are distributed across one or more \texttt{.mo} files so that a single file may contain several class definitions, and a single support class may be reused by many models. For example, a \texttt{Resistor} class in the Modelica standard library defines the electrical behavior of a resistor via Ohm's law, declares two connector pins, and can be reused by any circuit model that imports it.

\vspace{5pt}

\noindent\emph{Simulation-eligible classes.}
Model classes may include an \texttt{experiment} annotation specifying simulation parameters, such as start and stop times.
Such annotations indicate that the class is intended to be run as a standalone simulation scenario rather than only reused as a supporting class (e.g., a connector, partial model, or utility function).
We refer to classes carrying this annotation as \emph{simulation-eligible}.

\vspace{5pt}

\noindent\emph{Modelica compilation pipeline.}
Before a simulation-eligible class can be simulated, a Modelica compiler — the OpenModelica Compiler (OMC), for instance — builds a \textit{canonical representation}: the compiler loads packages, performs syntactic and type checking, elaborates the class hierarchy, strips semantically irrelevant annotations (e.g., visual/graphical annotations), and \emph{flattens} inheritance and connections into a system of equations ready for a numerical solver~\cite{fritzson2014principles}.
The result is a \emph{total model}: a single, large, self-contained \texttt{.mo} file in which all transitive dependencies have been inlined.

\vspace{5pt}

\noindent\emph{The Modelica standard library (MSL).}
The MSL is the reference open-source library distributed with Modelica tools.
It contains reusable components for mechanics, electrical, fluid, and thermal systems, as well as blocks for mathematics and other domains.
Because the MSL has been maintained for many years in a public repository, it provides both breadth (across many modeling domains and language constructs) and depth (a long evolutionary history), making it a natural target for the validation used in this paper.

\section{Related Datasets and Benchmarks}

CPS benchmarks typically evaluate tools rather than provide model corpora: \emph{AI-CPS} supplies Simulink plants and controllers checked against signal temporal logic specifications~\cite{DBLP:conf/icse/SongLZWZM22}, and \emph{CPSBench} annotates software-requirements documents with problem-frame entities to evaluate LLMs on CPS requirements modeling~\cite{DBLP:journals/corr/abs-2408-02450}.

Closest to ModBench in a neighboring modeling ecosystem is \emph{SLNET}, a corpus of Simulink models mined automatically from GitHub and MATLAB Central with project- and model-level metrics~\cite{DBLP:conf/msr/ShresthaCC22}.
SLNET captures single-point project snapshots, whereas ModBench mines commit histories, targets evolution, and validates the executability of each class snapshot by compiler.
Boll~\cite{boll2026data_desert} frames the broader Simulink ``data desert'' and complements SLNET with a search engine, an anonymizer for sharing industrial models, and a large-scale model synthesizer.
The Modelica ecosystem has no comparable corpus or tooling; ModBench is the first to mine Modelica library histories into compiler-validated, commit-traceable classes.

Prior Modelica work addresses library distribution, conformance testing, and compiler-side analysis rather than dataset construction. Tiller and Winkler's \emph{impact}~\cite{tiller2014impact} is a GitHub-centric package manager with a harvested JSON index and semantic versioning, but only flat, version-pinned resolution at library-release granularity.
The Modelica Association's \emph{ModelicaCompliance} suite~\cite{modelica_compliance} provides hand-curated specification test cases with expected outcomes for compiler conformance, not a mined corpus of evolving real-world models.
Fors et al.~\cite{fors2018safe} select regression tests safely in the OPTIMICA Compiler Toolkit using class-level static dependency rules verified by mutation testing on the MSL; the result is a dependency graph for test selection, not a persistent corpus of class snapshots.

Outside the CPS domain, empirical software engineering has long relied on benchmarks linking artifacts to historical data.
\emph{Defects4J}~\cite{DBLP:conf/issta/JustJE14}, for example, mines isolated bug-fixing commits from Java projects and pairs each with a bug-exposing developer test behind a uniform build abstraction.
ModBench adopts this mining-from-history paradigm with two adaptations: the unit of interest is a class snapshot rather than a bug-fix pair, and the executable oracle is compiler flattening rather than a developer-written test.

\section{The ModBench Pipeline}
\label{sec:pipeline}

\begin{figure*}
\centering
\includegraphics{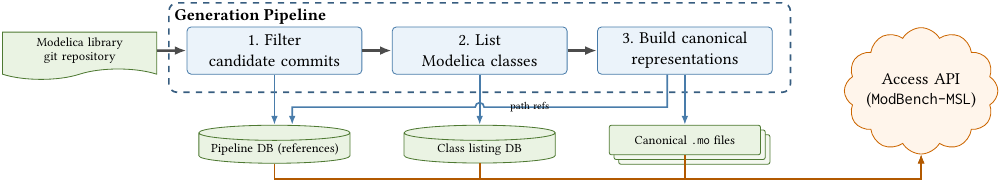}
\caption{Overview of ModBench. A configured Modelica library feeds
Steps~1--3 of the generation pipeline, which produce two databases and
a directory of canonical \texttt{.mo} files. The Access API provides a
unified read interface over all three artifact stores.}
\Description{Pipeline overview showing a Modelica library repository flowing through three pipeline steps: filter candidate commits, list Modelica classes, and build canonical representations. These steps produce two databases and a directory of canonical \texttt{.mo} files, all accessed through a common API.}
\label{fig:modbench-pipeline}
\end{figure*}

The ModBench generation pipeline (\autoref{fig:modbench-pipeline}) transforms a Modelica project from a Git repository into a dataset through three stages. First, candidate-commit filtering (\autoref{sec:pipeline-filtering}) selects which commits are included. Second, class listing (\autoref{sec:pipeline-listing}) traverses the selected commits to enumerate all Modelica classes. Third, canonical-representation building (\autoref{sec:pipeline-canonicalization}) compiles a total model for each simulation-eligible class. Each stage persists its results to database tables or an on-disk directory before the next stage begins.

\subsection{Filtering Candidate Commits}
\label{sec:pipeline-filtering}

The pipeline begins with a source configuration that specifies all information needed to process a Modelica project repository: the repository path, an ordered list of dependencies (package files to load), the default branch, branches to exclude, and a cutoff tag (or commit hash) defining the earliest commit to consider. This ensures the pipeline applies to any Modelica library with Git history.

The purpose of this stage is to reduce the repository history to candidate commits for class-level extraction by applying input restrictions (\textit{i.e.,} branches to exclude and cutoff tag) and content-based heuristics.
Concretely, the pipeline traverses the source repository's commit history (attributing file changes in merge commits to their first-parent integration path) and excludes commits matching any of the following four filters (applied in order), whose effect on the MSL is reported in \autoref{tab:filtering-funnel}:
(1)~\emph{Excluded branches:} commits reachable only from user-declared branches (typically maintenance or documentation branches), since their changes do not represent the main development trajectory of the library;
(2)~\emph{Pre-cutoff commits:} commits older than the configured cutoff tag or hash;
(3)~\emph{Bot-like commits:} commits whose author or message matches heuristic regex patterns for bot-generated commits, since they reflect tool actions rather than human modeling decisions;
(4)~\emph{No \texttt{.mo} files touched:} commits that do not modify any \texttt{.mo} file.

\begin{table}
\caption{Commit filtering (MSL).}
\label{tab:filtering-funnel}
\centering
\small
\begin{tabular}{lrr}
\toprule
Filter stage & Commits & Remaining \\
\midrule
Input unique commits & - & \MbFilterInputCommits{} \\
Excluded branches & \MbFilterExclBranches{} & \MbFilterAfterBranches{} \\
Before Modelica v3 & \MbFilterExclPreVThree{} & \MbFilterAfterPreVThree{} \\
Bot-like commits & \MbFilterExclBot{} & \MbFilterAfterBot{} \\
No touched \texttt{.mo} files & \MbFilterExclNoMo{} & \MbFilterRetained{} \\
\midrule
Retained candidates & \textbf{\MbFilterRetained{}} & \textbf{\MbFilterRetainedPct{}\%} \\
\bottomrule
\end{tabular}
\end{table}

\noindent\textbf{MSL instantiation.}
For the MSL instantiation, the history is restricted to commits after the \texttt{v3.0} tag to avoid mixing older language semantics with the modern library structure.
Despite this restriction, the retained history still spans nearly two decades.
In addition, documentation and legacy maintenance branches are excluded.
Restricting the file-touch filter to \texttt{.mo} files is a deliberate choice: external C source changes can affect simulation behavior without touching any \texttt{.mo} file, but this is not a concern for the MSL because OMC always uses its latest compiled external C support.
\autoref{tab:filtering-funnel} reports the resulting filtering, which reduces \MbFilterInputCommits{}~unique commits to \MbFilterRetained{}~retained candidates (\MbFilterRetainedPct{}\%).

\subsection{Listing Modelica Classes}
\label{sec:pipeline-listing}

ModBench treats Modelica classes as the primary dataset units because Modelica semantics is defined at the class level (rather than the file level) and simulation eligibility is a per-class property.
For each retained commit, the pipeline checks out the repository, loads the configured MSL packages in OMC (\MbToolOmc{}), and lists all classes in the loaded library hierarchy.
For each class, the listing records whether it carries an \texttt{experiment} annotation; this annotation is used in the next stage to select candidates for canonicalization.
The full class listing is recorded in a separate database; non-experiment classes are retained for potential use in future studies.

\noindent\textbf{MSL instantiation.} \autoref{tab:listing-statistics} reports validation metrics for this stage on MSL.
Most class versions are support artifacts such as connectors, records, functions, blocks, and partial models; only a small simulation-eligible subset carries an \texttt{experiment} annotation, with at most \MbListMaxExperimentPerCommit{}~such classes in a single commit.
In subsequent stages, the repository history is interpreted using the listed Modelica classes to derive meaningful benchmark units.

\begin{table}
\caption{Class listing statistics (MSL).}
\label{tab:listing-statistics}
\centering
\small
\begin{tabular}{lr}
\toprule
Quantity & Value \\
\midrule
Processed commits & \MbFilterRetained{} \\
Distinct class names & \MbListDistinctClassNames{} \\
All class versions & \MbListAllClassVersions{} \\
Non-simulation-eligible class versions & (\MbListNonExperimentPct{}\%) \MbListNonExperimentVersions{} \\
Simulation-eligible class versions & (\MbListExperimentPct{}\%) \MbListExperimentVersions{} \\
\bottomrule
\end{tabular}
\end{table}

\subsection{Building Canonical Representations}
\label{sec:pipeline-canonicalization}

This stage produces one canonical \texttt{.mo} file per simulation-eligible class version as defined in \autoref{sec:background-modelica}.
For this, ModBench invokes OMC to obtain a total model for the selected class.

From the full class listing produced in the previous stage, only classes carrying an \texttt{experiment} annotation are selected for canonicalization.
We use this annotation as a practical approximation for simulation eligibility: it may miss useful support-only classes or unannotated tests, but the retained classes meet an objectively verifiable validity criterion: they must simulate.
Extending canonicalization to any classes is left for future work.
Each canonical file is stored under a path that includes the source name and commit hash, and the database records the class name, path, compiler messages, and whether canonicalization succeeded.

For scalability, isolated worktrees and parallel workers process commits independently, avoiding checkout conflicts and enabling resumable runs from the last committed database state.
Canonical files are stored in source and commit keyed paths, and database entries record status and diagnostics for each extraction attempt.

\begin{table}
\caption{Canonicalization statistics (MSL).}
\label{tab:canonicalization-statistics}
\centering
\small
\begin{tabular}{lr}
\toprule
Quantity & Value \\
\midrule
Commits with $\geq$1 simulation-eligible class & (\MbCanonCommitsWithExperimentPct{}\%) \MbCanonCommitsWithExperiment{} \\
Simulation-eligible class versions (input) & \MbListExperimentVersions{} \\
Canonical class versions produced & (\MbCanonProducedPct{}\%) \MbCanonProducedVersions{} \\
Distinct canonical snapshots & 85{,}562 \\
Canonicalization failures & (\MbCanonFailuresPct{}\%) \MbCanonFailures{} \\
\bottomrule
\end{tabular}
\end{table}

\noindent\textbf{MSL instantiation.} To validate this stage at scale, the MSL instantiation produced \MbCanonProducedVersions{}~canonical class versions (\autoref{tab:canonicalization-statistics}). The \MbCanonFailures{}~recorded failures (\MbCanonFailuresPct{}\% of canonicalization attempts) mostly stem from historical library states incompatible with the single OMC version we use. \autoref{tab:failure-categories} groups them by the error patterns in OMC diagnostic: most are operator-record constructs rejected by stricter modern restriction checks introduced in later Modelica language versions, followed by references to classes renamed, moved, or not yet introduced in the checked-out commit.
Recording these failures rather than silently discarding them serves two purposes.
First, it makes the process auditable: users can distinguish absent records from records that were attempted but could not be canonicalized.
Second, it exposes realistic cases for Modelica tooling, since historical libraries contain transient states such as unfinished commits that differ from releases.
However, these results depend on the OMC services used; another compiler or version could resolve historical constructs differently and reclassify some failures.

\begin{table}
\caption{Canonicalization failure categories (MSL).}
\label{tab:failure-categories}
\centering
\small
\begin{tabular}{p{0.4\linewidth}rr}
\toprule
Failure family & Count & \% \\
\midrule
Operator-record restriction & \MbFailOperatorRestriction{} & \MbFailOperatorRestrictionPct{} \\
Missing referenced class & \MbFailMissingClass{} & \MbFailMissingClassPct{} \\
Replaceable/redeclare conflict & \MbFailRedeclare{} & \MbFailRedeclarePct{} \\
Missing base class & \MbFailMissingBase{} & \MbFailMissingBasePct{} \\
Canonicalization timeout & \MbFailTimeout{} & \MbFailTimeoutPct{} \\
Package load failure & \MbFailPackageLoad{} & \MbFailPackageLoadPct{} \\
\bottomrule
\end{tabular}
\end{table}

\section{The ModBench Dataset}
\label{sec:dataset}

Running the pipeline in \autoref{sec:pipeline} yields a dataset, shown in \autoref{fig:modbench-pipeline}, with three persistent stores: a pipeline database, a class-listing database, and canonical Modelica files.
\autoref{tab:schema-summary} summarizes their contents and roles, the statistics explained in \autoref{sec:dataset-statistics}, and the API described in \autoref{sec:api}.

\begin{table*}
\caption{Schema and artifact summary for the ModBench MSL dataset.}
\label{tab:schema-summary}
\centering
\small
\begin{tabular}{p{0.14\textwidth}p{0.1\textwidth}p{0.34\textwidth}p{0.23\textwidth}r}
\toprule
Store & Table & Main contents & Purpose & MSL size \\
\midrule
\multirow[t]{3}{*}{\begin{tabular}[t]{@{}p{0.11\textwidth}@{}}Pipeline DB\end{tabular}}
& \texttt{step1\_sources} & Source name, repository path, package load
targets, excluded branches, cutoff tag & Records the configuration needed for each repository to reproduce results.
& \multirow[t]{3}{*}{\vspace{0pt}$\approx$\MbStorePipelineDbSize{}} \\
& \texttt{step1\_commits}, \texttt{step1\_files} & Commit hash,
author/message metadata, exclusion reason, touched \texttt{.mo} files &
Preserves the filtered repository history and file-level traceability. & \\
& \texttt{step3\_classes}, \texttt{step3\_failures} & Source,
commit hash, class name, canonical path, production status; failure type
and compiler message & Links each experiment class version to its
canonical file or failure status. & \\
\midrule
Class listing\ DB
& \texttt{step2\_classes} & Source, commit hash, class name,
\texttt{is\_experiment} flag & Stores the full class listing.
& $\approx$\MbStoreClassEnumDbSize{} \\
\midrule
Canonical \texttt{.mo} files
& N/A & Standalone canonical Modelica
source per class & Stores the primary model artifacts.
& $\approx$\MbStoreCanonicalDirSize{} \\
\bottomrule
\end{tabular}
\end{table*}

\subsection{ModBench-MSL Dataset Statistics}
\label{sec:dataset-statistics}

As reported in the last column of \autoref{tab:schema-summary}, the class-listing store dominates the relational artifacts because it records every (commit, class) pair across \MbFilterRetained{}~commits and \MbListDistinctClassNames{}~distinct class names, whereas the \MbCanonProducedVersions{}~canonical \texttt{.mo} files from \autoref{sec:pipeline-canonicalization} account for most of the on-disk footprint.

\begin{figure}
\centering
\includegraphics[width=\columnwidth,keepaspectratio]{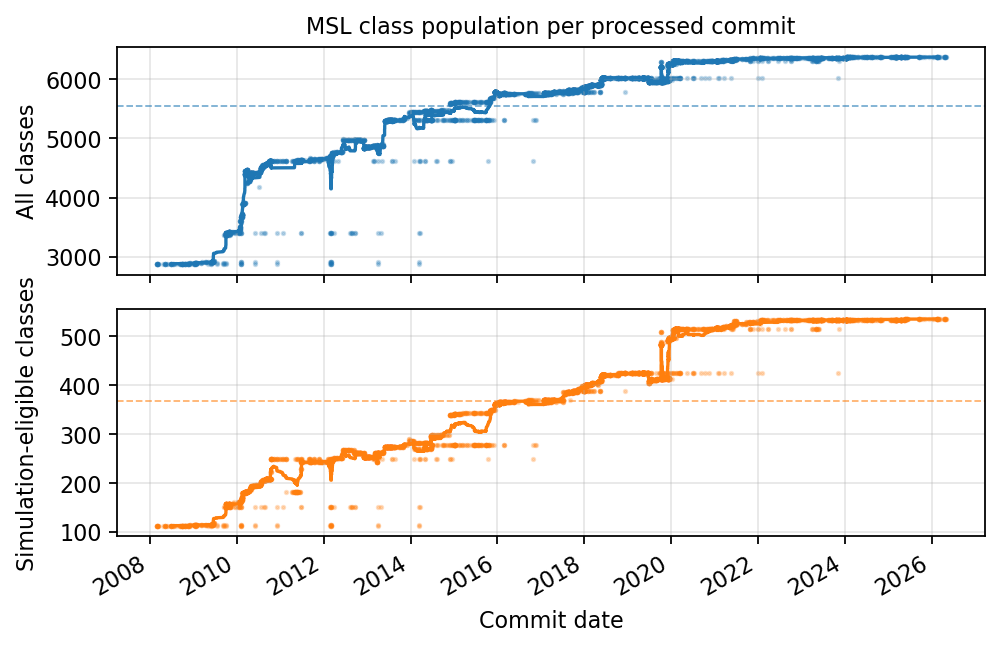}
\caption{MSL class counts at each processed commit over the project's history since Modelica v3. \emph{Top:} all classes loaded by OMC. \emph{Bottom:} simulation-eligible experiment classes only.}
\Description{Two time-series plots showing per-commit class counts across the MSL history since Modelica v3. The upper plot shows the total number of classes increasing from a few hundred to a peak of \MbGrowthAllClassMax{}. The lower plot shows simulation-eligible classes growing in parallel on a much smaller scale, peaking at \MbListMaxExperimentPerCommit{}.}
\label{fig:msl-class-timeline}
\end{figure}

Beyond aggregate counts, per-commit class populations show how the MSL has grown over time.
\autoref{fig:msl-class-timeline} plots the number of listed classes per processed commit over the project's history, split into all loaded classes and the simulation-eligible experiment subset.
All-class counts grow from a few hundred in the earliest retained commit to over 6{,}000 in recent commits (mean~\MbGrowthAllClassMean{} classes per commit), while the experiment subset grows in parallel but at a much smaller scale (mean~\MbGrowthExperimentMean{} classes per commit).
Plateaus and steps in both panels correspond to periods of stabilization and major version-bump expansions.
On average, only $\sim$\MbGrowthExperimentRatioMean{}\% of the classes loaded in a commit are simulation-eligible, and this ratio is stable across the history.
This indicates that simulation-eligible classes are a small but consistently maintained subset throughout the MSL's evolution, rather than a transient artifact of any particular release.

\subsection{Dataset Access and Distribution} \label{sec:api}

ModBench includes a read-only Python API that wraps the databases and canonical file tree.
The API supports listing source libraries, commits, and experiment classes; retrieving all canonicalized class versions in commit order; listing all models associated with a commit; reading canonical source for a specific class at a specific commit; and inspecting recorded canonicalization failures.
It also includes helpers for fetching GitHub commit, pull request, and issue metadata when such links are available.
When reading a canonical source, the API first checks the on-disk cache; if the canonical file is unavailable (e.g., due to its large size), it invokes OMC to regenerate the canonical representation for the requested class and commit.

The distribution is intended to be reproducible rather than merely a static dump.
The release package~\cite{modbench2026repo} includes generation scripts, configurations, SQLite databases, canonical Modelica files, and an exploratory notebook demonstrating common API queries.
This combination allows users to either use the prepared MSL dataset or run the same pipeline on their own Modelica repositories.

\section{Example Usages and Research Opportunities}

ModBench enables analyses that require both repository history and Modelica-aware semantic units: researchers can retrieve a model's timeline and inspect how its canonical representation evolves, while compiler developers can replay class snapshots across different OM versions.
We organize the broader opportunities below.

\emph{Model evolution.}
How artifacts change as a system matures is a long-standing question in software engineering, but has rarely been studied for equation-based models due to scarce per-class historical data.
By exposing the entire MSL history at canonical-class granularity, ModBench enables identification of stable versus frequently revised classes, detection of large-scale refactoring periods, and characterization of physical-component library lifecycles.
Insights into Modelica evolution can inform library design guidelines and help maintainers prioritize testing effort on volatile components.

\emph{Automated model repair and generation.}
Automated program repair and LLM-assisted code generation rely on benchmarks pairing real-world artifacts with bug-fixing or refactoring history, well established for mainstream languages (e.g., Defects4J~\cite{DBLP:conf/issta/JustJE14}) but largely absent for Modelica, limiting the transfer of these techniques to equation-based modeling~\cite{DBLP:conf/profes/SadrnezhaadLMV25}.
By retaining both successful canonical snapshots and non-compilable historical variants with their compiler diagnostics, ModBench provides realistic before-and-after pairs for evaluating Modelica-specific repair, bug localization, and code-generation tools on examples that reflect genuine engineering decisions rather than synthetic snippets.

\emph{Repository-linked empirical studies.}
Empirical studies of practitioner behavior, such as how bugs are reported, triaged, and fixed, depend on linking code artifacts to issue trackers and review discussions.
ModBench preserves a stable mapping from each canonical snapshot back to its commit, and the access API resolves these references to GitHub commit, pull request, and issue metadata.
This enables studies of, for example, the relationship between MSL refactorings and reported issues, the latency between bug introduction and fix, or the effect of community review on model quality.

\emph{Compiler regression testing.}
Modelica compilers evolve continuously, and changes to elaboration or flattening can silently break previously simulable models or alter their diagnostics.
Detecting such regressions is hard without a large, version-stamped corpus of real models~\cite{fors2018safe}.
ModBench's canonical, self-contained \texttt{.mo} files can be replayed across multiple compiler versions to surface regressions in successful compilations, observe how error messages drift over time, and measure improvements in front-end coverage.
This directly benefits compiler and downstream tool developers who need quantitative evidence of release-to-release behavior.

\section{Conclusion and Future Work}

We introduced ModBench, a general-purpose, reproducible pipeline for constructing Modelica benchmark datasets from library histories.
Our MSL validation demonstrates that class-level, compiler-assisted mining scales to thousands of commits and class snapshots while retaining traceability to original artifacts.
By combining filtered Git history, class listing, and canonical Modelica source for simulation-eligible classes, ModBench provides a foundation for empirical research on Modelica model evolution, tool evaluation, and automated model generation or repair.

Future work will extend the corpus by applying the pipeline to additional Modelica libraries and community repositories, as well as OMNotebook and Python-API Modelica code, to broaden coverage across modeling domains.

\begin{acks}
The first author was supported by the Vinnova competence center on Continuous Digitalization (CoDiG), while the last, partially, by the Wallenberg AI, Autonomous Systems and Software Program (WASP) funded by the Knut and Alice Wallenberg Foundation.
\end{acks}

\bibliographystyle{ACM-Reference-Format}
\bibliography{bibs}


\end{document}